# *Fourier-Based Spectral Analysis with Adaptive Resolution*

Andrey Khilko

email:  155tm2@gmail.com

**Abstract**

Despite being the most popular methods of data analysis, Fourier-based techniques suffer from the problem of static resolution that is currently believed to be a fundamental limitation of the Fourier Transform. Although alternative solutions overcome this limitation, none provide the simplicity, versatility, and convenience of the Fourier analysis. The lack of convenience often prevents these alternatives from replacing classical spectral methods – even in applications that suffer from the limitation of static resolution.

This work demonstrates that, contrary to the generally accepted belief, the Fourier Transform can be generalized to the case of adaptive resolution. The generalized transform provides backward compatibility with classical spectral techniques and introduces minimal computational overhead.

**Introduction**

The measurement of spectral characteristics of data involves a fundamental trade-off that is often referred to as the Uncertainty Principle: better resolution in frequency domain results in lower resolution in time domain and vice versa. In existing Fourier-based methods, resolution is generally static: decision about the time-frequency resolution trade-off must be made *a priori* and cannot be changed in process of measurement based on actual characteristics of measured data without prior knowledge of these characteristics. If processing of real time non-stationary signals is required, and their properties are not known in advance, existing Fourier-based analysis cannot provide accurate information about the presence of a particular frequency in the data spectrum and the time interval where the frequency is present. Due to a mismatch between the decided measurement resolution and the actual characteristics of the measured data causing undesirable smearing in both, time and frequency domains, the accuracy of such analysis is generally lower than the fundamental limit that is implied by the Uncertainty Principle. A similar problem exists when data contains concurrent spectral components with different uncertainties that cannot be analyzed jointly when resolution is static.

Although the time-frequency resolution trade-off is the most common manifestation, the Uncertainty Principle can also manifest itself as spatial resolution vs. frequency resolution. For clarity reasons, only the time-frequency resolution is mentioned below in this paper. In actual application, either manifestation can arise depending on the application nature.

A number of alternatives have been proposed to address the problem of static resolution of Fourier-based analysis. The most popular one is probably Multiresolution Analysis (MRA) based on Continuous Wavelet Transform (CWT) or Discrete Wavelet Transform (DWT; a good explanation of wavelets and MRA can be found e.g. in [1]). However, the CWT is computationally expensive and operates different terminology whereas the DWT provides nonlinear frequency grid where resolution is not easily scalable from practical prospective. Therefore in most cases, these alternatives cannot be immediately used as direct replacements of Fourier-based techniques and require new analysis methodology.

Because the Fourier Transform (FT) is historically by far the most popular engineering tool for analyzing characteristics of data, a need still exists for an enhanced Fourier-based method capable of working with non-stationary signals. However, the scientific community currently agrees in belief that the FT itself cannot provide this functionality. Therefore, when the FT needs to be used beyond the realm of static resolution, researches seek for application-specific workarounds.

In work [2], the FT was used for spectral analysis of speech signals. Because all phonemes[1] could not be reliably identified using same time-frequency resolution, the latter was tuned based on known phonetic structure of speech by adjusting length of sampling window for individual phonemes. While this approach can be applied to speech (known as a highly redundant signal) or sometimes even to a signal of unknown nature [3], it may not work well in situations when important spectral features that require different resolutions are concurrent and cannot be resolved serially.

---

[1] The smallest units of speech that distinguish meaning.

The problem of concurrent spectral components was addressed in [4]. Instead of varying sampling window length for different data fragments, the same data was analyzed multiple times using different static resolutions. One can see however, that this approach introduces serious computational expenses without adequate solution for the problem of general data processing. Applications reconstructing signal from processed data would require intermediate conversion back into time domain after processing at each particular resolution followed by FT with different length of sampling window.

A simpler method of signal reconstruction from multiple spectra of various time-frequency resolutions was suggested in [5]. Same data was analyzed and processed multiple times (using different static resolutions) and the results were mixed together. One can see however, that the procedure of mixing spectra with different time-frequency resolutions is application-dependent and may introduce artifacts.

This paper presents a novel technique to generalize the FT to the case of adaptive resolution. The proposed generalized transform provides capabilities of MRA without significant computational overhead and is backward compatible with the "classical" FT using complex exponents as base functions. This compatibility enables its proliferation as a general purpose tool in a majority of applications that rely on traditional spectral analysis, but suffer from a lack of MRA [6]. Spectral analysis based on the proposed transform can reach the fundamental limit implied by the Uncertainty Principle for a bounded set of time-frequency resolutions rather than for a fixed predefined value.

## 1. The Idea of Adaptive Resolution in Spectral Analysis

A sequence of *2N* samples can be analyzed as a single sequence or as two sequences of *N* samples each. The former approach gives us one spectrum with *2N* frequencies and time resolution $2N/f_s$ (where $f_s$ is sampling rate) whereas the latter yields two sequential spectra with only *N* frequencies, but better time resolution $N/f_s$. Because the signal bandwidth in both cases is the same (according to the Sampling Theorem, it only depends on sampling rate) the count of frequencies directly translates into frequency resolution (continuous range of frequencies covered by single frequency component in discrete spectrum) that will be $\pi/N$ and $2\pi/N$ correspondingly. Thus at the given sampling rate, the length of the sampling sequence is the sole variable that controls the time vs. frequency resolution tradeoff. Until now, there has been no known way to change this variable once spectrum is obtained. If a different resolution was required, one had to calculate the corresponding spectrum from scratch. The absence of a known way to transform resolution of taken spectra has been interpreted as the impossibility of such a transform. A closer examination of this problem reveals that one can obtain some additional information from joint analysis of two consecutive spectra by using the fact that the spectra are consecutive.

Short Time FT of *2N* samples for sequence $X_n$ starting from sample $n_t$, where *k* is frequency index is as follows:

$$x(k, n_t, 2N) = \sum_{n=n_t}^{n_t+2N-1} X_n W_{2N}^{kn}, \quad \text{where} \quad W_Z^Y = \exp\left(\frac{-2\pi i Y}{Z}\right); \quad k = 0, \ldots, 2N-1.$$

The sum of *2N* elements can be split into two sums of *N*. Then, substituting variable in the second sum:

$$x(k, n_t, 2N) = \sum_{n=n_t}^{n_t+2N-1} X_n W_{2N}^{kn} =$$

$$\sum_{n=n_t}^{n_t+N-1} X_n W_{2N}^{kn} + \sum_{n=n_t+N}^{n_t+2N-1} X_n W_{2N}^{kn} = \sum_{n=n_t}^{n_t+N-1} X_n W_{2N}^{kn} + \sum_{n'=n_t}^{n_t+N-1} X_{n'} W_{2N}^{k(n'+N)}; \quad n = n'+N.$$

Since $W_Z^{X+Y} = W_Z^X W_Z^Y$ and $W_{2N}^{kN} = W_2^k = \exp(-\pi i k) = (-1)^k$,

$$x(k,n_t,2N) = \ldots = \sum_{n=n_t}^{n_t+N-1} X_n W_{2N}^{kn} + \sum_{n'=n_t}^{n_t+N-1} X_{n'} W_{2N}^{k(n'+N)} =$$

$$\sum_{n=n_t}^{n_t+N-1} X_n W_{2N}^{kn} + W_{2N}^{kN} \sum_{n'=n_t}^{n_t+N-1} X_{n'} W_{2N}^{kn'} = \sum_{n=n_t}^{n_t+N-1} X_n W_{2N}^{kn} + (-1)^k \sum_{n'=n_t}^{n_t+N-1} X_{n'} W_{2N}^{kn'}.$$

Now let's substitute variable *k* with *k'*:   *k* = 2*k'* for even *k* and *k* = 2*k'*+1 when *k* is odd -

$$x(2k', n_t, 2N) = \sum_{n=n_t}^{n_t+N-1} X_n W_{2N}^{2k'n} + (-1)^{2k'} \sum_{n'=n_t}^{n_t+N-1} X_{n'} W_{2N}^{2k'n'} =$$

$$\sum_{n=n_t}^{n_t+N-1} X_n W_N^{k'n} + \sum_{n'=n_t}^{n_t+N-1} X_{n'} W_N^{k'n'} = x(k', n_t, N) + x(k', n_t + N, N); \quad k' = 0, \ldots, N-1.$$

Thus, the even frequencies in the *2N* - sample spectrum can be expressed in terms of frequencies obtained from two consecutive *N* - sample spectra.

The odd frequencies in the *2N* - sample spectrum cannot be directly obtained from the *N* - sample spectra. This would require frequencies that are not normally calculated as part of the Discrete FT (they are not needed to recreate the original sequence by inverse transform):

$$x(2k'+1, n_t, 2N) = \sum_{n=n_t}^{n_t+N-1} X_n W_{2N}^{(2k'+1)n} + (-1)^{2k'+1} \sum_{n'=n_t}^{n_t+N-1} X_{n'} W_{2N}^{(2k'+1)n'} =$$

$$\sum_{n=n_t}^{n_t+N-1} X_n W_N^{(k'+1/2)n} - \sum_{n'=n_t}^{n_t+N-1} X_{n'} W_N^{(k'+1/2)n'} = x(k'+1/2, n_t, N) - x(k'+1/2, n_t + N, N).$$

However, if the *N* - sample spectra $x^R$ contain these additional redundant frequencies then two such consecutive spectra can be easily transformed into single spectrum with twice the frequency resolution and half the resolution in time domain:

$$x^R(k, n_t, N) = \sum_{n=n_t}^{n_t+N-1} X_n W_{2N}^{kn}; \quad k = 0, \ldots, 2N-1 \qquad (1)$$

$$x(k, n_t, 2N) = x^R(k, n_t, N) + (-1)^k x^R(k, n_t + N, N).$$

When spectrum with redundant frequencies $x^R$ is used by itself the redundancy should be discarded, but it is the redundancy that allows subsequent resolution transformation.

Because the both spectra, classical *x* and redundant $x^R$, are based on the same sampling rate they cover the same bandwidth and relationship between their corresponding frequencies is given by the following expression:

$$x(k, n_t, N) = x^R(2k, n_t, N) \ ; \quad k = 0, \ldots, N-1.$$

The even frequencies of the redundant spectrum $x^R$ correspond to the frequencies of the classical spectrum $x$ and the additional redundant frequencies are located in between them:

$$x(k + 1/2, n_t, N) = x^R(2k+1, n_t, N) \ ; \quad k = 0, \ldots, N-1.$$

So, we have demonstrated by (1) how the time-frequency resolution can be transformed by factor 2 when the initial spectra contain certain redundancy. In the next section, the resolution transform is presented for general case of a natural factor.

The idea of the proposed resolution transform is to take a number of initial consecutive spectra with high resolution in time domain (short sequence of samples) and then to improve their frequency resolution by combining two or more consecutive spectra (effectively increasing length of analyzed sampling sequence). Because such operation increases count of resolved frequencies, the initial spectra must provide redundant frequencies allowing subsequent transformation into higher spectral resolutions. Therefore, length of sampling sequence to calculate the initial spectrum determines the highest resolution in time domain whereas redundancy factor – maximal possible improvement of frequency resolution when consecutive redundant spectra are combined together.

## 2. Operators of Redundant Spectral Transform and Resolution Transform

If count $N^0$ is not prime it can be represented as a product of at least two natural numbers $N_j$:

$$N^0 = \prod_{j=0}^{H-1} N_j \ ; \quad H \geq 2.$$

To generalize the above expression, $N^x$ will be partial product of these multipliers:

$$N^x = \prod_{j=x}^{H-1} N_j \ ; \quad 0 \leq x < H.$$

It may look confusing at first sight, but the convention is simple: index in lower corner means a multiplier, and index in upper corner – a product of one or more multipliers.

Indices $n^x$ and $n_x$ will be running from zero and up to $N^x$ -1 and $N_x$ -1 correspondingly:

$$n^x = 0, \ldots, N^x - 1; \quad n_x = 0, \ldots, N_x - 1; \quad 0 \leq x < H.$$

Since $N_{H-1}$ is the last multiplier in the product, it is obvious that

$$N^{H-1} = N_{H-1}; \quad n^{H-1} = n_{H-1}. \qquad (2)$$

Now, we are ready to go. Let's assume that $N^0$ is count of samples in a sampling sequence to be analyzed. Each its sample can be denoted by index $n^0$. The critical point for understanding this paper is that because the $N^0$ count is not prime, single index $n^0$ can be substituted with a vector of indices $[n_0, n_1, \ldots, n_{H-1}]$ by using the following recurrence relation:

$$n^j = n^{j+1} N_j + n_j; \quad j = 0, \ldots, H - 2. \qquad (3)$$

When using this relation iteratively, each iteration increases number of dimensions by one. For example, instead of denoting a specific sample by index $n^0$ we can logically split all samples into $N^1$ groups of $N_0$ samples each and use index of the group $n^1$ and index of the sample within the group $n_0$. In the next iteration, index $n^1$ will be substituted with $n^2$ and $n_1$, etc. until index $n_{H-1}$ is reached (Fig.1). Thus, instead of using index $n^0$ we can use the following vector:

$$n^0 = n^1 N_0 + n_0 = (n^2 N_1 + n_1)N_0 + n_0 = n^2 N_1 N_0 + n_1 N_0 + n_0 =$$
$$(n^3 N_2 + n_2)N_1 N_0 + n_1 N_0 + n_0 = n^3 N_2 N_1 N_0 + n_2 N_1 N_0 + n_1 N_0 + n_0 = \ldots = N^0 \sum_{j=0}^{H-1} \frac{n_j}{N^j} = \bar{n}.$$

What will happen to FT of the complete sequence of $N^0$ samples when the recurrence relation (3) is iteratively applied to it?

$$f(k) = \sum_{n^0=0}^{N^0-1} X_{n^0} W_{N^0}^{k n^0}; \quad k = 0, \ldots, N^0 - 1.$$

As a first iteration, we substitute $n^0$ with $(n^1 N_0 + n_0)$ and use the fact that the base function is exponent so, $W_Z^{X+Y} = W_Z^X W_Z^Y$ and $W_{YZ}^{XY} = W_Z^X$:

$$f(k) = \sum_{n^0=0}^{N^0-1} X_{n^0} W_{N^0}^{kn^0} = \sum_{n^1 N_0+n_0=0}^{N^1 N_0-1} X_{n^1 N_0+n_0} W_{N^0}^{k(n^1 N_0+n_0)} = \sum_{n^1=0}^{N^1-1}\sum_{n_0=0}^{N_0-1} X_{n^1 N_0+n_0} W_{N^0}^{kn^1 N_0} W_{N^0}^{kn_0} = \sum_{n^1=0}^{N^1-1}\sum_{n_0=0}^{N_0-1} X_{n^1 N_0+n_0} W_{N^1}^{kn^1} W_{N^0}^{kn_0}.$$

One can see that the middle part of the expression under the sums does not depend on the inner index $n_0$, so it can be moved outside of the inner sum:

$$f(k) = \ldots = \sum_{n^1=0}^{N^1-1}\sum_{n_0=0}^{N_0-1} X_{n^1 N_0+n_0} W_{N^1}^{kn^1} W_{N^0}^{kn_0} = \sum_{n^1=0}^{N^1-1} W_{N^1}^{kn^1} \sum_{n_0=0}^{N_0-1} X_{n^1 N_0+n_0} W_{N^0}^{kn_0}.$$

The inner sum in the above expression now looks very similar to FT of the sequence of $N_0$ samples. However, instead of $N_0$ frequencies it provides $N^0$ of them. We already met similar transform in section 1. By analogy with (1) one can conclude that it represents the FT with additional redundant frequencies. The only difference from (1) is that at this time there are $N^1$ additional frequencies between each two frequencies corresponding to the "classical" FT.

Thus, the definition of the Redundant Spectral Transform converting a sequence of $N$ samples into a redundant spectrum with redundancy factor $M$ is as follows:

$$f_j^0(k, N, M) = \sum_{n=0}^{N-1} X_{jN+n} W_{NM}^{kn}; \quad j = 0, \ldots, M-1; \quad k = 0, \ldots, NM-1. \tag{4A}$$

Index $j$ demonstrates that $M$ consecutive spectra with same resolution and redundancy factor are obtained. Parameter $k$ represents frequency index. When $M=1$, this transform degenerates into the classical FT.

After rewriting the FT of the sequence of $N^0$ samples using (4A),

$$f(k) = \sum_{n^0=0}^{N^0-1} X_{n^0} W_{N^0}^{kn^0} = \ldots = \sum_{n^1=0}^{N^1-1} W_{N^1}^{kn^1} \sum_{n_0=0}^{N_0-1} X_{n^1 N_0+n_0} W_{N^0}^{kn_0} = \sum_{n^1=0}^{N^1-1} W_{N^1}^{kn^1} f_{n^1}^0(k, N_0, N^1)$$

one can see that it is expressed in terms of frequencies obtained from $N^1$ consecutive redundant spectra $f^0(k, N_0, N^1)$.

In second iteration, expression (3) is applied to index $n^1$:

$$f(k) = \ldots = \sum_{n^1=0}^{N^1-1} W_{N^1}^{kn^1} f_{n^1}^0(k, N_0, N^1) = \sum_{n^2=0}^{N^2-1} W_{N^2}^{kn^2} \sum_{n_1=0}^{N_1-1} W_{N^1}^{kn_1} f_{n^2 N_1+n_1}^0(k, N_0, N^1).$$

Instead of transforming all $N^1$ spectra $f^0$ directly into classical Fourier spectrum $f(k)$ one can logically split $N^1$ spectra into $N^2$ groups of $N_1$ each, and perform the Resolution Transform in two steps: first within the groups then between the groups.

After comparing the results of the first two iterations it becomes clear that the following operator transforms resolution of a sequence of redundant spectra to produce a single spectrum with lower redundancy and higher resolution in frequency domain:

$$f_j^{i+1}(k, NL, M/L) = \sum_{n=0}^{L-1} W_M^{kn} f_{jL+n}^i(k, N, M); \quad k = 0, \ldots, NM - 1. \tag{5A}$$

When a sequence of $L$ redundant $N$-sample spectra is combined into a single spectrum the latter covers $L$ times more samples than each individual one. Therefore, frequency resolution of the resulting spectrum is enhanced by factor $L$. Note, that this operation also reduces redundancy factor $M$ and time resolution by $L$ (so, $M$ must be divisible by $L$). When all redundancy is transformed ($M=L$), the result is equal to FT of the complete sequence of $NM$ samples.

Rewriting the result of the second iteration with the help of (5A)

$$f(k) = \ldots = \sum_{n^2=0}^{N^2-1} W_{N^2}^{kn^2} \sum_{n_1=0}^{N_1-1} W_{N^1}^{kn_1} f_{n^2 N_1 + n_1}^0(k, N_0, N^1) = \sum_{n^2=0}^{N^2-1} W_{N^2}^{kn^2} f_{n^2}^1(k, N_0 N_1, N^2)$$

and using (3) and (5A) over and over again until index $n_{H-1}$ is reached,

$$f(k) = \ldots = \sum_{n^2=0}^{N^2-1} W_{N^2}^{kn^2} f_{n^2}^1(k, N_0 N_1, N^2) = \ldots = \sum_{n^{H-1}=0}^{N^{H-1}-1} W_{N^{H-1}}^{kn^{H-1}} f_{n^{H-1}}^{H-2}(k, N_0 N_1 \cdots N_{H-2}, N^{H-1})$$

then applying (2),

$$f(k) = \ldots = \sum_{n^{H-1}=0}^{N^{H-1}-1} W_{N^{H-1}}^{kn^{H-1}} f_{n^{H-1}}^{H-2}(k, N_0 N_1 \cdots N_{H-2}, N^{H-1}) =$$

$$\sum_{n_{H-1}=0}^{N_{H-1}-1} W_{N^{H-1}}^{kn_{H-1}} f_{n_{H-1}}^{H-2}(k, N_0 N_1 \cdots N_{H-2}, N_{H-1}) = f_0^{H-1}(k, N^0, 1)$$

one can see that the Resolution Transform (5A), when iteratively applied to redundant spectra, can provide a set of various time-frequency resolutions of the same data. The initial spectrum $f^0$ has the worst frequency resolution $2\pi/N_0$ and the best time resolution $N_0/f_s$, the final spectrum $f^{H-1}$ - the best frequency resolution $2\pi/N^0$ and the worst time resolution $N^0/f_s$, and an intermediate iteration $x - (2\pi N^{x+1}/N^0)$ and $(N^0/(N^{x+1}f_s))$ correspondingly. Count of possible iterations is determined by set of selected multipliers of $N^0$. These multipliers are not necessarily prime. Therefore, there may be more than one way to reach a particular resolution. Because the end result does not depend on a chosen way to split $N^0$ into natural multipliers, the iteration index does not provide significant information and may be omitted:

$$f_j^X(k,N,M) = f_j^Y(k,N,M) = f_j(k,N,M).$$

Thus, we have demonstrated that the FT of a sequence of $N^0$ samples can be expressed in terms of iterative transforms gradually increasing resolution in frequency, but lowering that in time domain:

$$f(k) = f_0^{H-1}(k,N^0,1) = \sum_{n_{H-1}=0}^{N_{H-1}-1} W_{N^{H-1}}^{kn_{H-1}} \sum_{n_{H-2}=0}^{N_{H-2}-1} W_{N^{H-2}}^{kn_{H-2}} \cdots \sum_{n_0=0}^{N_0-1} W_{N^0}^{kn_0} X_{\bar{n}} = U_{N^{H-1}}^{N_{H-1}} U_{N^{H-2}}^{N_{H-2}} \cdots U_{N^1}^{N_1} R_{N^0}^{N_0} X_{\bar{n}}$$

where $R$ is the operator of Redundant Spectral Transform

$$R_{NM}^N = \sum_{n=0}^{N-1} W_{NM}^{kn}; \quad k = 0, \ldots, NM-1; \quad R_{NM}^N : X_n \to f(k,N,M) \tag{4B}$$

and $U$ – the operator of Resolution Transform

$$U_M^L = \sum_{n=0}^{L-1} W_M^{kn}; \quad k = 0, \ldots, NM-1; \quad U_M^L : f_n(k,N,M) \to f(k,NL,M/L). \tag{5B}$$

Since the multipliers $N_j$ do not have to be prime and their order may be arbitrary,

$$U_{M/L_2}^{L_1} U_M^{L_2} = U_{M/L_1}^{L_2} U_M^{L_1} = U_M^{L_1 L_2}.$$

As a result of multiple resolution transforming iterations (5B) one can obtain a plurality of equivalent data representations with different time-frequency resolutions. Each representation can be used individually as a classical Short Time Fourier spectrum after the redundant frequencies are discarded. Therefore, the transforms (4B) and (5B) provide capability of MRA being fully compatible with the Short Time FT methodology. An example of MRA for a sequence of 16 samples is shown schematically in Fig.2.

The expressions (4A) and (4B) can be formally extended to the case of continuous transform. Although being technically trivial, the transition from sums to integrals does not add clarity to the idea of adaptive resolution. So, the author leaves it as an optional exercise for the reader.

## 3. Normalization Conventions

Normalization factors and the sign in the exponent can be treated in the same way as in the classical FT: they can be conventionally moved between the forward and the inverse transforms. However for joint analysis of multiple resolutions, it may be convenient to have the normalization factor $1/L$ multiplying the Resolution Transform, so that the inverse FT for all resolutions has same normalization and the results of all resolution iterations can be directly compared to each other:

$$f_p(k, NL, M/L) = \frac{1}{L} \sum_{j=0}^{L-1} f_{j+pL}(k, N, M) W_M^{kj}; \quad L > 0; p = 0, \ldots, M/L - 1. \tag{5C}$$

If the results of different measurements with different initial resolutions need to be compared then it may be convenient to place the normalization factor $1/N$ in front of the Redundant Spectral Transform:

$$f_j(k, N, M) = \frac{1}{N} \sum_{n=0}^{N-1} X_{n+jN} W_{NM}^{kn}; \quad M, N > 0; j = 0, \ldots, M-1; k = 0, \ldots, NM - 1. \tag{4C}$$

These are the normalization conventions followed in the rest of the paper.

## 4. Inverse Transforms of Redundant Spectrum

The inverse transform (6A) shows an example of how to discard redundancy to use a spectrum with redundancy factor $M$ in place of a Short Time Fourier spectrum. One can see that only one out of each $M$ consecutive frequencies is used. The same method of skipping (discarding) the redundant frequencies can be used in any application where backward compatibility with classical spectrum is required. When $M=1$ (6A) degenerates into classical inverse FT.

$$X_{n+jN} = \sum_{k=0}^{N-1} f_j(kM, N, M) W_N^{-kMn}; \quad j = 0, \ldots, M-1; n = 0, \ldots, N-1. \tag{6A}$$

Index $j$ demonstrates that there are $M$ consecutive spectra with same resolution and redundancy factor.

It is easy to prove that another way to approach inverse transform of redundant spectrum may be to include the redundancy in the transform rather than discarding it:

$$X_{n+jN} = \frac{1}{M} \sum_{k=0}^{NM-1} f_j(k, N, M) W_{NM}^{-kn}. \tag{6B}$$

However, additional research is required to investigate whether (6B) provides any advantages over (6A) that could justify extra computation associated with it.

It is remarkable that the resolution transforming iterations (5C) can be applied to a partial set of frequencies, so different frequency ranges can be independently transformed to different resolutions. The only constraint is if the inverse transform is needed then all frequencies involved in it must have same resolution.

## 5. Iterational Improvement of Frequency Resolution

Fig. 4 demonstrates how close frequencies can be gradually resolved in real time by enhancing frequency resolution in spectral analyzer. The analyzed data is a simulated signal comprised from three close frequencies (Fig. 3). Dashed curve in Fig. 4 is initial spectrum *f (k, 32, 8)* calculated from sequence of first 32 samples with redundancy factor *M*=8. This spectrum cannot resolve individual contributing frequencies due to the Uncertainty Principle. However after receiving next 32 samples, the frequency resolution can be doubled by combining information from two consecutive spectra with *M*=8 into spectrum *f (k, 64, 4)* (dotted curve), so that the second peak is resolved. Finally, the third spectrum *f (k, 128, 2)* that resolves all three peaks, is obtained after next 64 samples are received and two consecutive spectra *f (k, 64, 4)* are combined.

One can see that, unlike DWT, the transforms (4C) and (5C) provide real time MRA i.e. in order to obtain a spectrum with high resolution in time domain from partial sequence one does not need to wait for the complete sequence to be received. Spectral resolution can be improved gradually upon receiving sufficient count of samples. When spectral analyzer has adequate computational power to operate on continuous flow of samples, latency of a spectrum with a particular resolution does not depend on maximal frequency resolution the analyzer is configured to.

Gradual transformation of resolution may be useful when signal timing is unknown. To avoid smearing of signal features either in time or frequency domain, it may be important to match position and length of sampling window to onset and duration of the measured signal. These characteristics may be estimated from a sequence of initial spectra with sufficient time resolution, so that higher frequency resolution with minimal smearing can be obtained by applying the Resolution Transform only to a subset of the initial spectra overlapping with the actual signal.

A similar idea implemented by a two-pass spectral analysis was discussed in [2], where first pass with fixed sampling window was used to adjust window of second pass. However despite of ideological similarity, the present technique allows achieving the same result by resolution transforming iterations rather than by separate passes. These iterations also provide benefit of sequential processing at various resolutions without intermediate conversion into time domain.

This approach may also help when frequency to be measured deviates over time. In such case, attempt to improve precision by enhancing frequency resolution may in fact deteriorate it if deviation over longer measurement period becomes significant. Therefore for optimal results, a trade-off between better frequency resolution and smaller amount of deviation may need to be established. For signals with unknown deviation rate, decision about the optimal frequency resolution may be made by analyzing the results of initial (4C) or intermediate (5C) resolution transforming iterations. E.g., they may be stopped when the result of the most recent iteration indicates significant difference in the measured frequency between consecutive spectra of same resolution.

## 6. Joint Analysis of Multiple Resolutions

Use case shown in Fig. 5 and Fig. 6 demonstrates a more sophisticated scenario of analyzing a composite signal containing both short pulses and continuous tones. In Fourier-based analysis with static time-frequency resolution, precise timing of the pulses and precise frequencies of the continuous tones cannot be measured jointly. But it can be done by MRA technique in two steps using (4C) and (5C) respectively. Before using these transforms, one needs to decide what should be maximal resolutions in time and frequency domains. The former determines parameter $N$ in (4C) (length of sampling sequence for initial spectra; it is assumed that sufficient sampling rate is technically feasible) and the latter – parameter M (redundancy factor; may be limited by performance of the analyzer). As a first step, a set of consecutive initial spectra with the maximal resolution in time domain is obtained (Fig. 5). These spectra provide required timing characteristics of the analyzed signal. The second step is a transformation of the redundancy into higher frequency resolution after sufficient count of initial spectra is collected (Fig. 6). One can see that joint analysis of the results of the two described steps provides accurate information about the both, timing of the pulses and frequencies of the continuous tones. When needed, the transformation of resolution may be gradual, generating in a single measurement a plurality of jointly analyzable data representations with various time-frequency resolutions. Changes introduced to a redundant spectrum before the Resolution Transform (5C) will be automatically carried over by (5C) into higher frequency resolution providing an easy generic method to superimpose processing at multiple time-frequency resolutions without intermediate conversion back into time domain or application-dependent mixing of spectra with different resolutions [5].

    An example of one interesting candidate for such application is modeling of human hearing. It is known, that at frequencies around 3-5 kHz (the area of the highest sensitivity) its resolution in time domain is better than 50 ms which corresponds (in classical spectral analysis) to frequency resolution being worse than 20 Hz. In fact, its actual frequency resolution in this spectral area is about an order of magnitude better. Even though the mechanism of human hearing seems to be different from (5C) (not all phase information is preserved) the set of transforms (4C) and (5C) provides good starting point, because loss of phase information can be easily simulated.

## 7. Nonrectangular Windowing

In certain real time applications, a continuous stream of samples is split into fragments, each fragment is individually processed, and the original signal is restored piece-by-piece by inverse transform from the processed spectra of sequential fragments. This procedure introduces distortions known as the spectral leakage. At given sampling rate, the leakage increases when the fragments become shorter, so an obvious way to deal with it is to select the fragment length that would be sufficiently large.

In classical Fourier analysis, the length of fragment may be constrained by required time-frequency resolution. Therefore in a real time application, it is not usually an available degree of freedom and one may need to deal with the introduced artifacts. This problem is typically worked around by careful selection of nonrectangular windowing function. In MRA, frequency resolution (determined by length of complete fragment which corresponds to value $NM$ in the above definitions) and time resolution (determined by length of partial fragment to obtain initial spectra and corresponding to value $N$) do not constitute a tradeoff, and the length of complete fragment can be sufficiently large. This approach may eliminate the need for nonrectangular windowing.

Unfortunately, the length of complete fragment may still be limited by maximal acceptable processing latency or computational complexity. In such case, nonrectangular windowing may be needed at the boundaries of a complete fragment. "Internal" boundaries between partial fragments may still be formed by rectangular window, because (5C) seamlessly merges them without adding to the spectral leakage.

However, rectangular edges at the "internal" boundaries inside the fragments may preclude one from using an intermediate redundant spectra to recreate the original signal in applications where adaptive response time is required. So in some cases, nonrectangular windowing may be needed for each partial fragment.

## 8. Redundant Spectrum of a Monochromatic Signal

Redundant spectrum of a monochromatic signal looks complex. It demonstrates periodic fringes caused by timing localization of the data (Fig. 7).

Analytical representation of spectrum of a complex-valued monochromatic signal with redundancy factor *M* obtained from sequence of *N* samples is

$$f(k_0, N, M) = \frac{1}{N}\sum_{n=0}^{N-1} W_{NM}^{-k_0 n} W_{NM}^{kn} = \frac{1}{N}\sum_{n=0}^{N-1} W_{NM}^{(k-k_0)n} \; ; \quad k = 0,\ldots, NM-1 \; ; \; 0 \le k_0 < NM \quad (7)$$

where $k_0$ – signal frequency and $k$ represents frequency index in signal spectrum.

If a monochromatic signal is real-valued, calculation of its spectrum by (7) can be optimized because spectrum of real-valued data is symmetric: as in the classical FT, it contains pairs of frequencies that are complex conjugates of each other. Therefore, it is possible to present only half of such spectrum limited by the critical frequency $NM/2$:

$$f(k_0, N, M) = \frac{1}{N}\sum_{n=0}^{N-1} \cos\left(\frac{2\pi k_0 n}{NM}\right) W_{NM}^{kn} =$$

$$\frac{1}{N}\sum_{n=0}^{N-1}\left[\frac{1}{2}W_{NM}^{-k_0 n} + \frac{1}{2}W_{NM}^{k_0 n}\right] W_{NM}^{kn} \; ; \quad k = 0,\ldots, NM-1 \; ; \; 0 < k_0 < \frac{NM}{2}. \quad (8A)$$

If the critical frequency $k_{crit} = NM/2$ exists ($NM$ is divisible by 2) then its spectral representation is

$$f(k_{crit}, N, M) = \frac{1}{N}\sum_{n=0}^{N-1} W_{NM}^{-nNM/2} W_{NM}^{kn} = \frac{1}{N}\sum_{n=0}^{N-1} e^{\pi i n} W_{NM}^{kn} = \frac{1}{N}\sum_{n=0}^{N-1}(-1)^n W_{NM}^{kn}. \quad (8B)$$

For frequency with index 0 (offset or "DC" component) corresponding representation is

$$f(0, N, M) = \frac{1}{N}\sum_{n=0}^{N-1} W_{NM}^{0n} W_{NM}^{kn} = \frac{1}{N}\sum_{n=0}^{N-1} W_{NM}^{kn}. \quad (8C)$$

## 9. Multiresolution Synthesis

The MRA principle presented in this work can be used in signal synthesis. Even though it is the inverse problem of spectral decomposition, the sequence of actions is not reversed. To obtain a spectrum representing concurrent components with different time-frequency resolutions, (5C) is to be used for resolution conversion in the same way as in spectral decomposition. Components with the highest localization in time domain generated by (7) or (8) are added together into synthesized initial redundant spectra. Then, after redundancy of the initial spectra is transformed into higher frequency resolution by (5C), components with higher localization in frequency domain are superimposed. The second step can be repeated multiple times until all redundancy is transformed. After that, the final spectrum containing all components is inverse Fourier transformed to provide the required composite signal in time domain. If a spectrum for inverse transform still contains redundancy, (6A) or (6B) should be used instead of the inverse FT. When two spectra are being added together their resolution and redundancy factors must be same. Redundancy factor of the initial synthesized spectra must be sufficient to achieve the highest intended frequency resolution. The result may be subject to nonrectangular windowing as described above in section 7.

## 10. Computational Complexity

To optimize transform (4C) for computation it may be convenient to rearrange its *NM* frequencies into *M* interleaved sets. It can be easily done by substitution of variable *k* with *k'* so that $k = k'M + m$:

$$f_j(k,N,M) = \frac{1}{N}\sum_{n=0}^{N-1} X_{n+jN} W_{NM}^{kn} = \frac{1}{N}\sum_{n=0}^{N-1} X_{n+jN} W_{NM}^{(k'M+m)n} = \frac{1}{N}\sum_{n=0}^{N-1} (X_{n+jN} W_{NM}^{mn}) W_{N}^{k'n} \quad ;$$

$$j = 0\ldots M-1 \quad ; \quad k = 0\ldots NM-1 \quad ; \quad k = k'M+m \quad ; \quad m = 0\ldots M-1 \quad ; \quad k' = 0\ldots N-1$$

One can see that as a result of this variable substitution (4C) gets decomposed into *M* separate transforms of redefined sequence $X^m$ :

$$f_j(k,N,M) = \frac{1}{N}\sum_{n=0}^{N-1} X_{n+jN} W_{NM}^{kn} = \cdots = \frac{1}{N}\sum_{n=0}^{N-1} X^m_{n+jN} W_{N}^{k'n} \quad ; \quad X_{n+jN} W_{NM}^{mn} = X^m_{n+jN} \quad . \quad (9)$$

When *m*=0 the original sequence remains unchanged ($X = X^0$) and the respective transform represents the "classical" FT of the original sequence *X*. When *m* > 0 the corresponding transform can be considered as another "classical" FT of redefined sequence $X^m$ obtained by multiplying each element of the original sequence by exponential factor. The frequencies obtained from the *M* transforms must be interleaved (see Fig.8). Because each transform may be calculated as FFT, computational complexity of (4C) is $O(MN \log_2 N)$. To obtain $M$ sequential redundant spectra to cover the complete sequence of $NM$ samples $O(M^2 N \log_2 N)$ operations are required because (4C) needs to be applied $M$ times.

Resolution Transform (5C) obviously decomposes into $NM$ independent transforms, one for each frequency (Fig.9). Regardless of whether all redundancy $M$ for a specific frequency is transformed in a single or in multiple iterations, $O(M)$ operations are required. So, total computational complexity of MRA based on (4C) and (5C) is

$$O(M^2 N \log_2 N) + O(M^2 N) = O(M^2 N \log_2(2N)) . \tag{10A}$$

When redundancy factor $M$ is fixed (fixed dynamic range of covered resolutions) the complexity exhibits asymptotic behavior similar to that of FFT. When $N$ is fixed and $M$ increases (fixed initial resolution, expanding dynamic range) the complexity behavior becomes quadratic.

The more expensive quadratic dependence on $M$ can be improved when no changes to intermediate resolution iterations are required. In such a case, computation of a particular redundant frequency can be postponed until the frequency stops being redundant (becomes resolved) as a result of the Resolution Transform. Then, FFT technique can be used to fully calculate the missing "postponed" frequencies without use of (5C). To estimate computational complexity of MRA with postponed redundancy, we will consider two extreme cases: single and maximal count of resolution transforming iterations.

In the case of a single Resolution Transform, it gets $O(MN \log_2 N)$ operations to obtain $M$ initial non-redundant spectra, $O(MN)$ operations to transform $N$ their frequencies by (5C) and $O(MN \log_2(MN))$ operations to calculate the missing postponed frequencies by using FFT technique. Therefore, computational complexity of MRA with single Resolution Transform and postponed redundancy is

$$O(MN \log_2 N) + O(MN) + O(MN \log_2(MN)) = O(MN \log_2(2MN^2)) . \tag{10B}$$

To maximize count of iterations for the other extreme case, each iteration should be increasing frequency resolution by the smallest possible factor. Because the smallest possible factor is 2, it is convenient to consider the total count of samples to be a power of 2 ($NM = 2^R$). Then, initial spectra will be obtained from 2 samples and each iteration will increase frequency resolution by factor 2. Because of postponed redundancy, spectra will be missing half of their frequencies after each Resolution Transform. These missing postponed frequencies will correspond to odd frequencies of "classical" FT and could be obtained by half of operations required by corresponding FFT algorithm (Fig.10). The other half (even frequencies) will be obtained by (5C) from results of previous iteration. After last iteration, there will be a single spectrum with $2^R$ frequencies half of which will be obtained by FFT in $O(1/2 \cdot 2^R \log_2 2^R) = O(2^{R-1} R)$ operations and the other half in $O(2^R)$. The previous iteration will have 2 spectra with $2^{R-1}$ frequencies, then 4 spectra with $2^{R-2}$ frequencies, etc down to the initial $2^{R-1}$ spectra consisting of only 2 frequencies. Then, total computational complexity will be

$$O\left(2^{R-1} \cdot 2 + \sum_{x=2}^{R} 2^{R-x}\left(2^{x-1}x + 2^x\right)\right) = O\left(2^R\left(1 + \sum_{x=2}^{R}\left(\frac{x}{2}+1\right)\right)\right) = O(2^R R^2) = O(MN \log_2^2(MN)) \,. \text{(10C)}$$

The schema of MRA with postponed redundancy becomes attractive when optimal time-frequency resolution for a particular fragment of data is estimated from preliminary analysis of this data (e.g. speech recognition). In such an application, required resolution for each fragment may be determined by analysis of initial spectra with high resolution in time domain. If these initial spectra already contain redundancy, some redundancy computation is "wasted" for the fragments that do not need maximal intended frequency resolution. On the other hand, attempt to minimize this "wasted" computation by lowering redundancy factor imposes limit on maximal frequency resolution. The both problems may be avoided by the postponed redundancy technique. When the initial spectra are not redundant, precise amount of redundancy may be added later to each spectrum eliminating the need of discarding excessive redundancy. Multiple resolution transforming iterations with postponed redundancy can be made until changes need to be introduced. Before introducing changes, maximal frequency resolution needs to be determined and corresponding amount of redundant frequencies must be inserted in spectra by using the FFT decomposition principle (9) demonstrated in Fig.8. Then, all changes will be carried over by (5C) to spectra with higher frequency resolution.

In deterministic systems designed to sustain the worst case scenario (to be able to process all data at given range of resolutions) or in systems where data transforms are implemented in hardware, performance optimization provided by the technique with postponed redundancy may be undesirable due to increased complexity of the transform algorithm. Such applications may choose to implement the transforms (4C) and (5C) in a straightforward way.

## Conclusions

The proposed Fourier-based MRA seems to be more convenient than wavelet-based as a tool for general use. Thanks to backward compatibility with the FT, it allows to preserve and reuse intellectual investments in existing spectral methodologies by extending them into domain of adaptive resolution and non-stationary signals. Unlike DWT (where increasing density of frequency grid cannot be achieved by simple adjustment of a transform parameter and requires designing a different filter bank), it is easily scalable and real time capable. In the proposed Fourier-based MRA, spectral resolution can be improved immediately upon receiving sufficient count of samples regardless of maximal intended frequency resolution. The wavelet-based MRA may, however, remain as the preferred analysis method in several situations, e.g. when exponent is not a convenient base function to represent original signal, or when higher computational efficiency of logarithmic frequency division is crucial.

## Acknowledgements

The author expresses gratitude to his wife, Natalia Mislavsky for encouraging support.

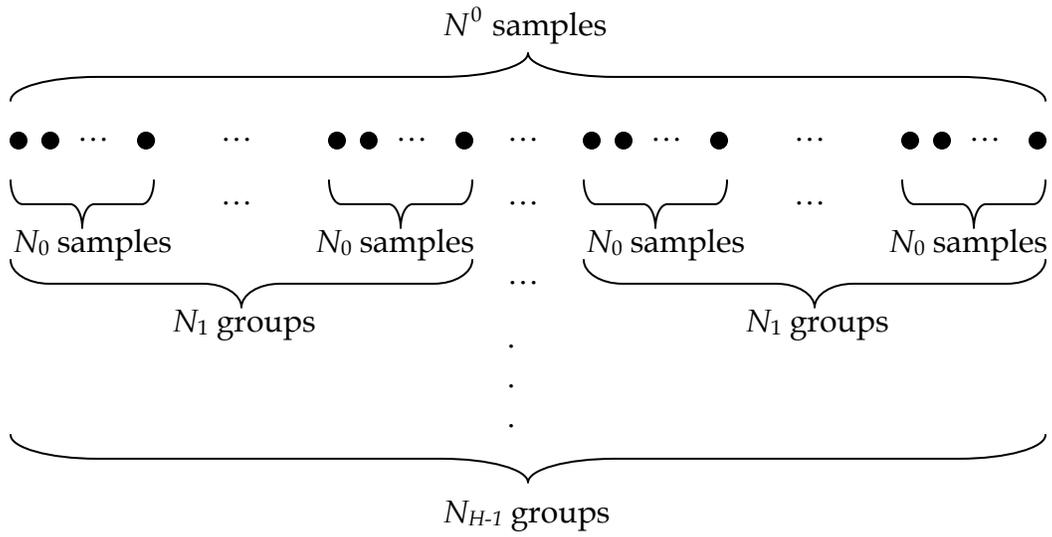

Fig.1 When $N^0$ is not prime, single index ($n^0$ - one dimension) can be substituted with vector of indices ($[n_0, n_1, …, n_{H-1}]$ - $H$ dimensions).

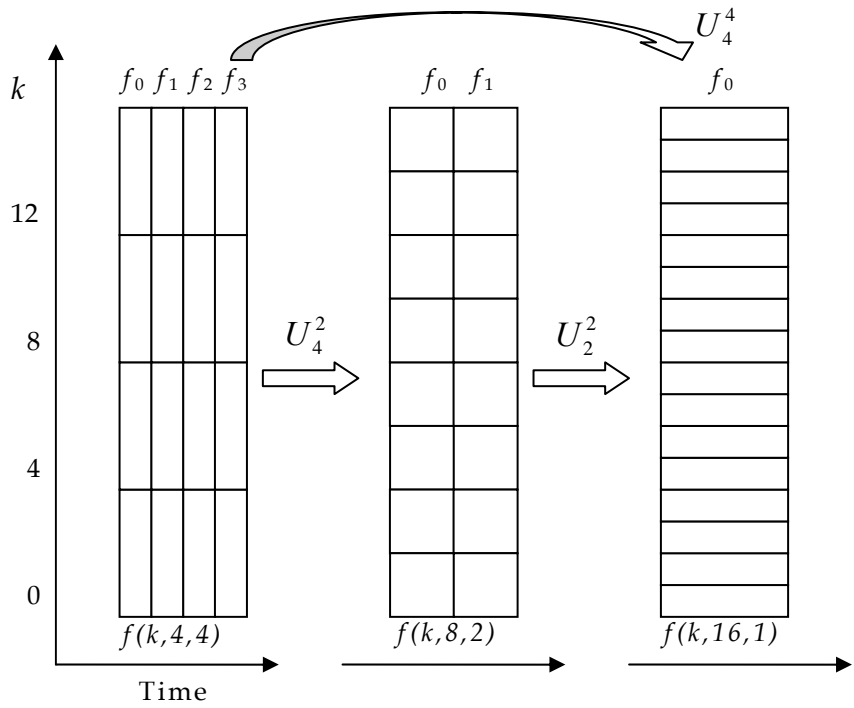

Fig.2 MRA schema for sequence of 16 samples with initial spectra obtained from 4 samples. Possible transitions between resolution cells.

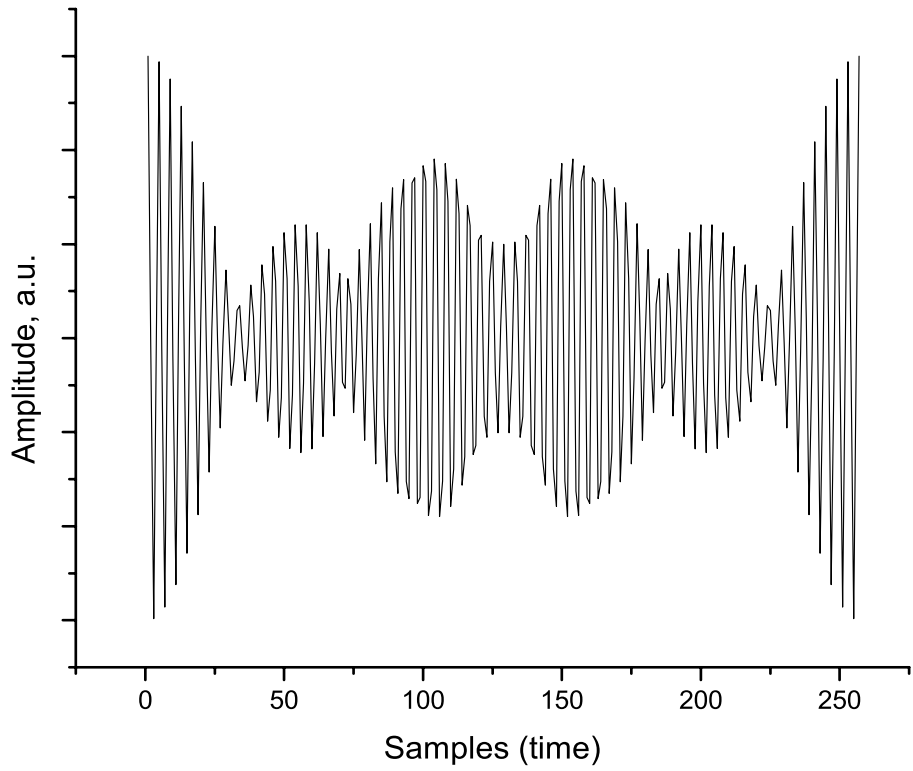

Fig. 3 is 256 samples of a simulated signal comprised from three concurrent continuous frequencies with close values around 0.25 (normalized by sampling frequency).

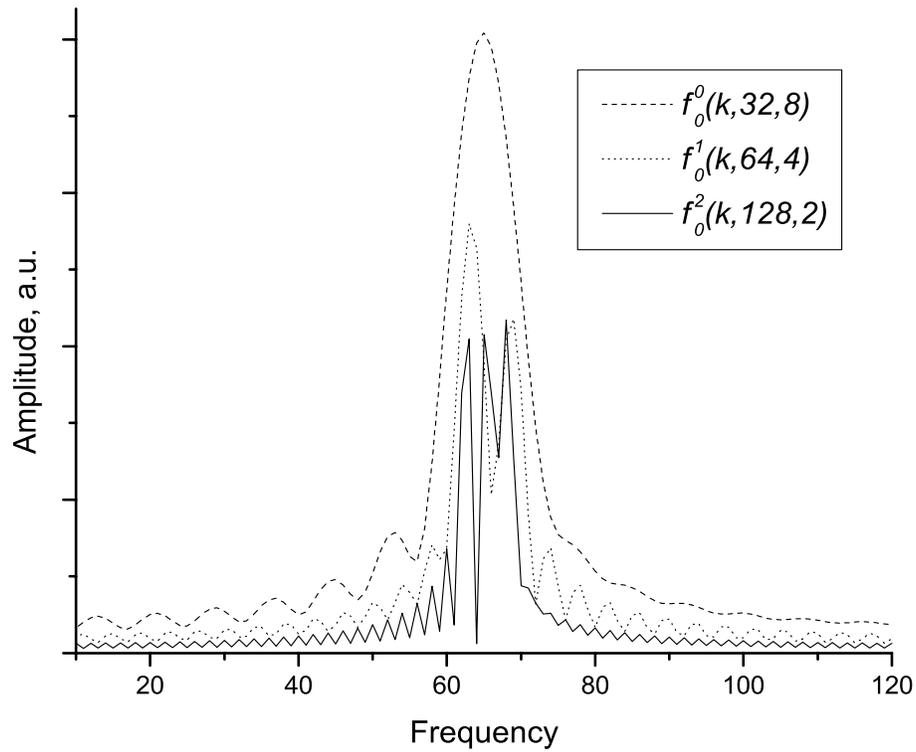

Fig. 4 presents three spectra with different resolutions obtained from the sampling sequence in Fig. 3: initial spectrum $f^0$ calculated over 32 samples with redundancy factor 8 (the dashed curve) and two subsequent iterations $f^1$ and $f^2$ each enhancing frequency resolution by factor 2 (the dotted and the solid curves respectively). The iterative process gradually resolves all three frequencies comprising the original signal in Fig. 3.

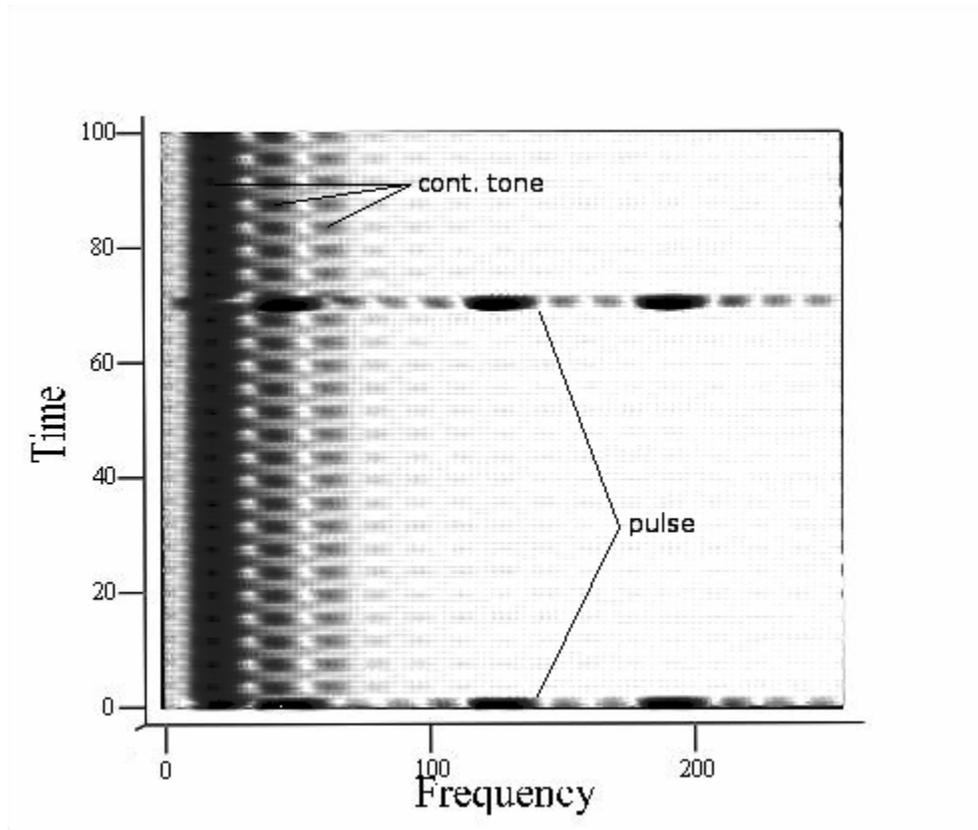

Fig. 5 demonstrates a spectrogram obtained from a simulated composite signal by using (4C). It displays two short pulses with good timing localization and three continuous tones with frequencies that cannot be precisely determined here due to the Uncertainty Principle. The spectrogram consists of a sequence of redundant spectra *f (k, 32, 16)*. Time should be interpreted as spectrum sequence index. The frequency #256 has normalized value 0.5 (the critical frequency).

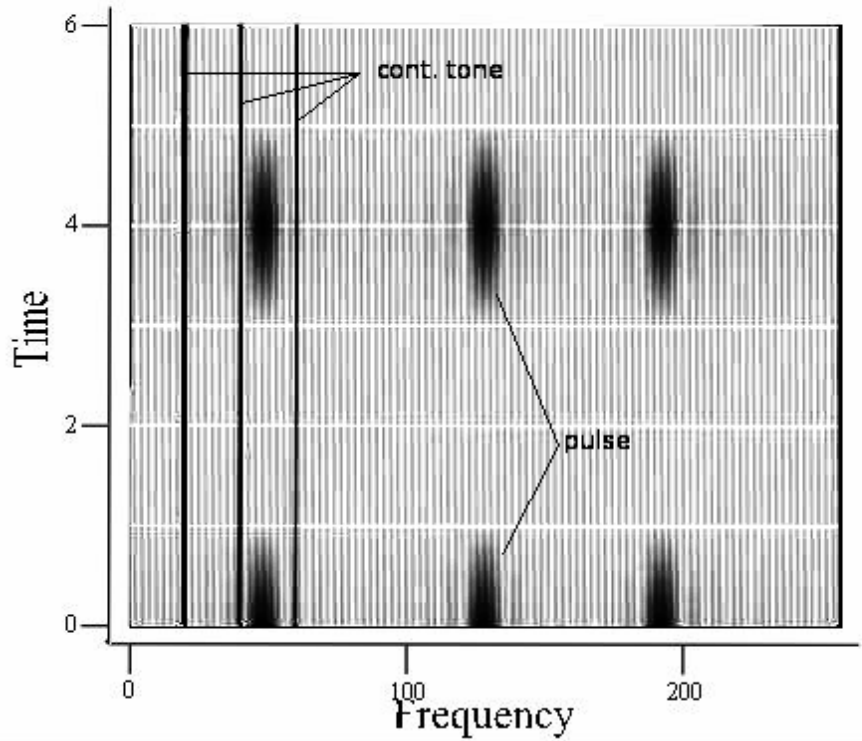

Fig. 6 demonstrates spectrogram corresponding to maximal frequency resolution *f (k, 512, 1)* obtained from the spectral data presented in Fig. 5 by using (5C). Frequencies of the three continuous tones can be measured precisely. Due to the Uncertainty Principle, precision of time measurements is lost. However, more precise timing characteristics of the pulses can be obtained from the spectrogram in Fig. 5. Because higher frequency resolution requires longer sampling sequences to be analyzed, 6 consecutive spectra in Fig. 6 cover same time interval as 96 consecutive spectra in Fig. 5.

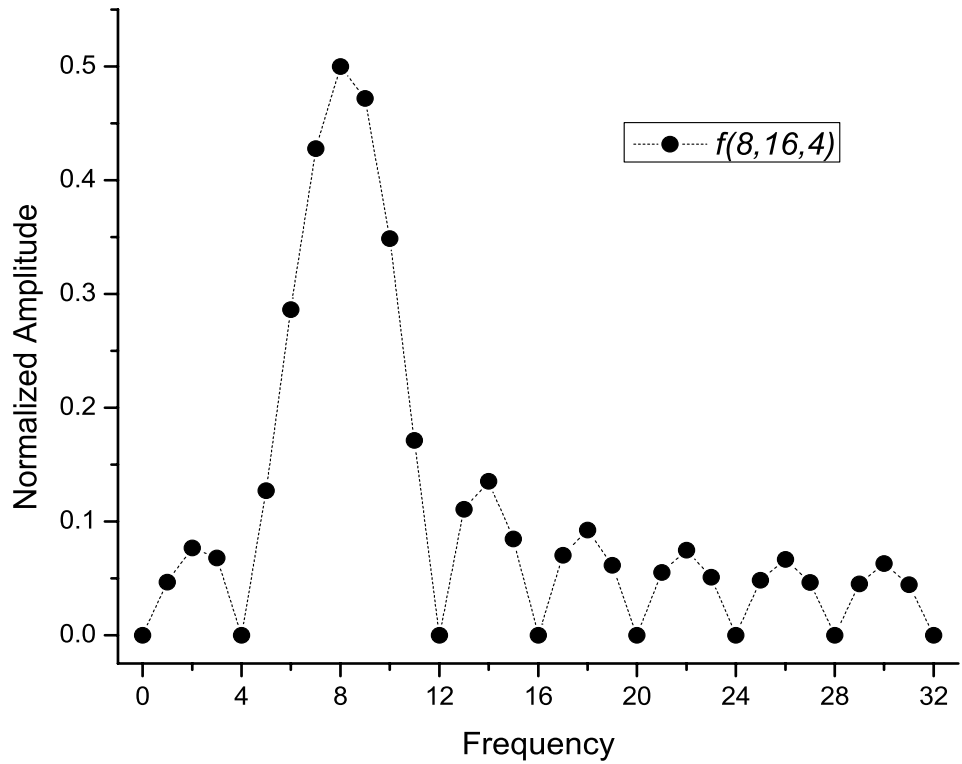

Fig. 7 is spectral representation of a real-valued monochromatic signal with frequency $k_0 = 8$ and redundancy factor 4 corresponding to 16 data samples. The critical frequency has index 32. This curve is plotted using (8A). Symmetric part of the spectrum is not shown.

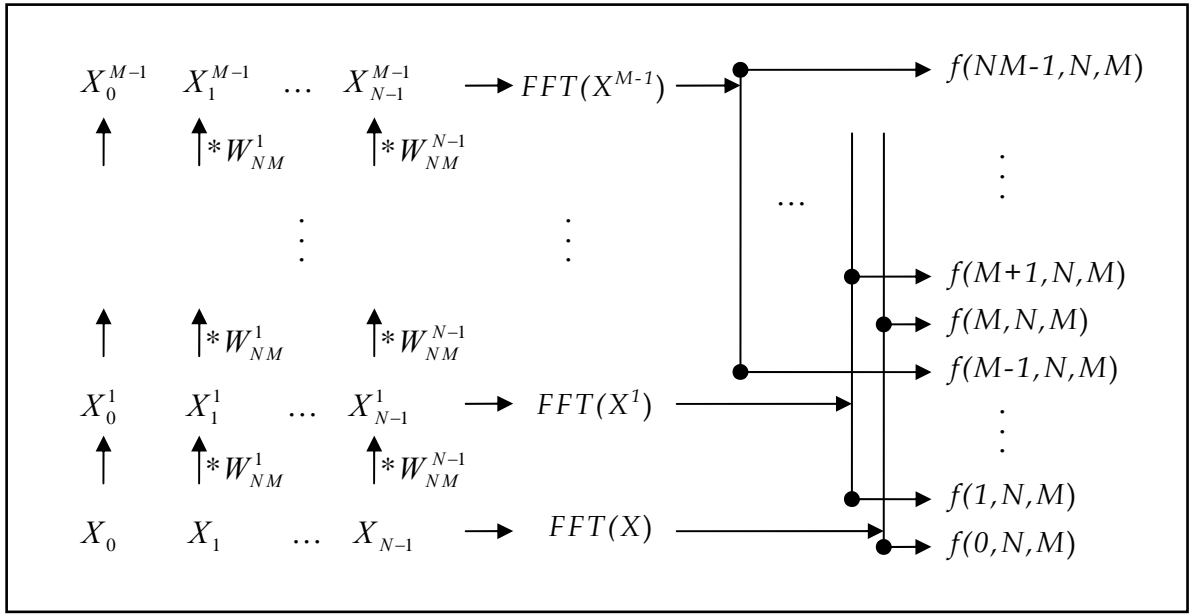

Fig.8 Decomposition of Redundant Spectral Transform with redundancy factor $M$ into $M$ interleaved Fourier Transforms.

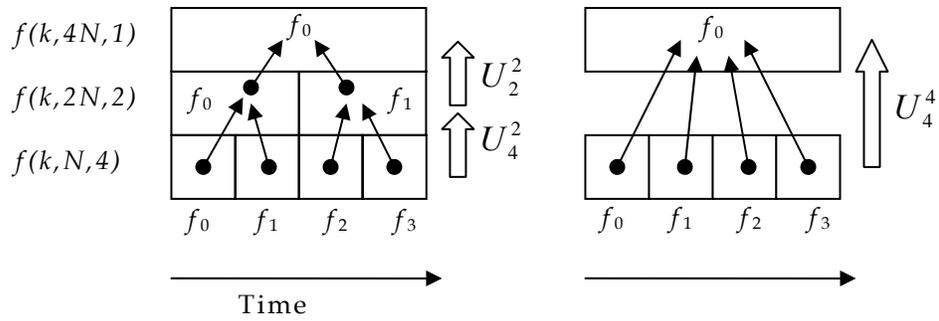

Fig.9 Example of decomposition of Resolution Transform into individual transforms for each frequency: multiple iterations on the left and single iteration on the right. Redundancy factor 4 is assumed.

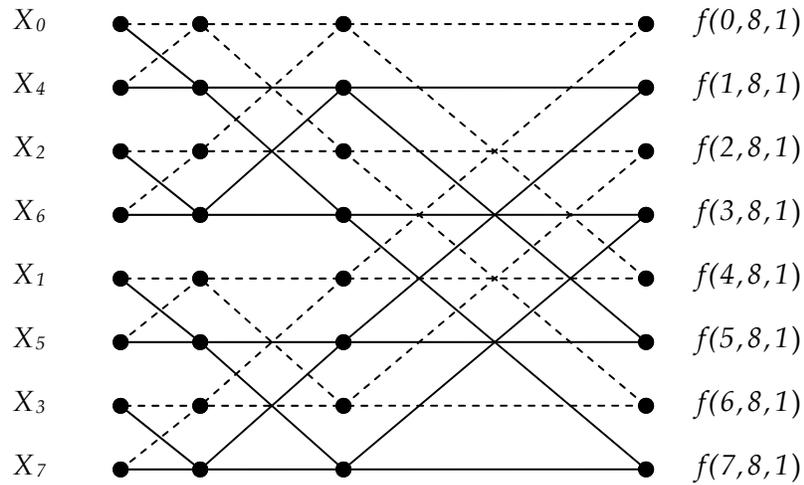

Fig.10 Example of FFT "butterfly" diagram for 8 samples. The algorithm can be decomposed into two computational flows providing even (dashed lines) and odd frequencies (solid lines). Therefore, only half of operations (comparing to full FFT) are required to obtain just odd frequencies.